\documentclass[aps,prl,twocolumn,floatfix,showpacs, superscriptaddress,]{revtex4-2}
\usepackage{amssymb,amsmath,amsfonts}
\usepackage{graphicx,graphics}
\usepackage{epsfig}
\usepackage{epstopdf}
\usepackage{amsfonts}
\usepackage{bm,bbm}
\usepackage{amssymb,amsmath,amsfonts}
\usepackage{mathrsfs}
\usepackage{calc}
\usepackage{epsfig}
\usepackage{float}
\usepackage{color}
\usepackage{epstopdf}
\usepackage{bm,amssymb,amsmath}
\usepackage{graphicx,color}

\usepackage{ulem}

\begin{document}

\title{A many-body Szil\'ard engine with giant number fluctuations}
\author{Omer Chor}
\author{Amir Sohachi}
\affiliation{Raymond \& Beverly Sackler School of Physics and Astronomy, Tel Aviv University, Tel Aviv 6997801, Israel}
\author{R\'emi Goerlich}
\author{Eran Rosen}
\affiliation{Raymond \& Beverly Sackler School of Chemistry, Tel Aviv University, Tel Aviv 6997801, Israel}
\author{Saar Rahav}
\email{rahavs@ch.technion.ac.il}
\affiliation{Schulich Faculty of Chemistry, Technion--Israel Institute of Technology, Haifa 3200008, Israel}
\author{Yael Roichman}
\email{roichman@tauex.tau.ac.il}
\affiliation{Raymond \& Beverly Sackler School of Physics and Astronomy, Tel Aviv University, Tel Aviv 6997801, Israel}
\affiliation{Raymond \& Beverly Sackler School of Chemistry, Tel Aviv University, Tel Aviv 6997801, Israel}

\date{\today}
\begin{abstract}
Szil\'ard's information engine is a canonical example in the field of thermodynamics of information. We realize experimentally a macroscopic many-particle Szil\'ard engine that consists of active particles and use it to lift a mass against gravity. We show that the extractable work per cycle increases when the raised weight is changed more gradually during the process. Interestingly, we find that the ideal extractable work grows with the number of particles due to giant number fluctuations. This is in contrast to the calculated behavior of a similar engine operating on thermal particles.
\end{abstract}

\maketitle

\begin{figure}
    \centering
    \includegraphics{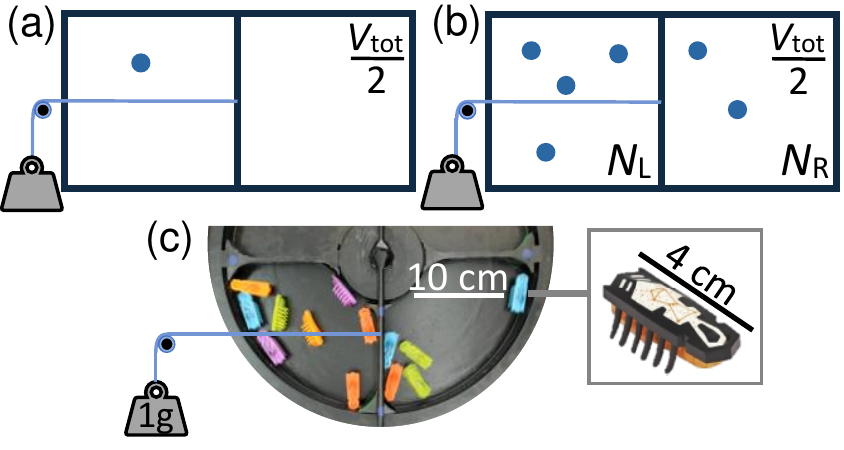}
    \caption{(\textbf{a}) Szil\'ard's engine. (\textbf{b}) Many-body generalization.  (\textbf{c}) Our experimental setup with many active particles. The arena is round and the partition is free to rotate on an axle placed in its center. A mass is attached to a string that is connected to the partition, converting the circular motion of the partition to vertical motion and pulling the mass against gravity (hence extracting work).}
    \label{fig:exp_setup}
\end{figure}

In recent years there has been a renewed interest in the connection between information and thermodynamics \cite{parrondo2015thermodynamics, landauer1961irreversibility, bennett1973logical, sagawa2008second, cao2009thermodynamics, horowitz2010nonequilibrium, sagawa2010generalized, ponmurugan2010generalized, fujitani2010jarzynski, kundu2012nonequilibrium}. This fundamental line of research, originating with Maxwell's celebrated thought experiment \cite{maxwell1871theory} is now a reality, having been experimentally tested in various settings \cite{admon2018experimental, jun2014high, berut2012experimental, toyabe2010experimental, ribezzi2019large, paneru2018lossless, saha2021maximizing, paneru2020colloidal, koski2014mutual, koski2014experimental, masuyama2018information, cottet2017observing, vidrighin2016photonic}. The first illustration of how a Maxwell's demon can be used to construct an engine is due to Szil\'ard \cite{szilard1929entropieverminderung}, who proposed a single-molecule engine directly converting information to work (Fig.~\ref{fig:exp_setup}a).

In Szil\'ard's engine, an ideal gas molecule is placed in a rectangular box that is in equilibrium with a surrounding heat bath. A mobile partition is then placed in the center of the box, and a measurement is made to determine which side the molecule occupies. The information gained by the measurement, $I=\ln2$, is proportional to the reduction in entropy of the system. Next, the single-molecule gas is allowed to expand isothermally, performing work of $W=k_BT\ln2$. The partition is then removed, bringing the system to its original state. As in any typical engine, this process is repeated in a cyclic manner.

While previous experimental realizations have mainly focused on single-particle systems, theoretical analyses have also explored generalizations to many-body information engines \cite{kim2011information, horowitz2011designing, song2021optimal, kim2011quantum, bengtsson2018quantum, cai2012multiparticle, jeon2016optimal}. In a classical many-body Szil\'ard's engine \cite{kim2011information} (Fig.~\ref{fig:exp_setup}b), $N$ ideal-gas particles are placed in the box, and the measurement determines the number of particles in each side, $N_L, N_R$ (for simplicity, an odd $N=N_L+N_R$ is assumed). Work is extracted from the system by taking advantage of the pressure difference between the two halves. In the many-body case, work extraction increases with the imbalance in the initial division of particles. This is the case for any system where the pressure is a monotonic function of the density (see SI section 1 \cite{comment0}).
 
Information engines are inherently out-of-equilibrium due to the application of measurement outcome-dependent feedback. In most of the literature, this is the sole mechanism that drives the engine away from equilibrium. Only very recently, connection to a nonequilibrium environment \cite{saha2023information, paneru2022colossal}, or an active working substance \cite{malgaretti2022szilard}, were considered.

Here, we consider a Szil\'ard engine operating on a many-body active system. In a Szil\'ard engine operating on an ideal gas, the number of particles in each half is distributed binomially, due to thermally induced number (or density) fluctuations. In contrast, if the engine operated instead on a system that exhibits giant number fluctuations, the imbalance between both halves would be more prominent. Giant number fluctuations are known to exist in driven dissipative systems \cite{Goldhirsch93, Narayan2007,deseigne2010collective}  and active matter \cite{Ramaswamy2003,toner2005hydrodynamics,ginelli2010large, chate2008collective,zhang2010collective, Palacci2013,bricard2013emergence}. Therefore, realizations with large number differences, $\left| N_L-N_R\right|$, are more likely to occur in a Szil\'ard engine with active particles. This, in turn, should result in a higher amount of extractable work. 

In the following, we describe our experimental realization of a many-body Szilard engine consisting of self-propelled particles (Fig.~\ref{fig:exp_setup}c) (see SI Movie 1 \cite{comment0}). We start by characterizing the properties of the active system, focusing on the distribution of $N_R$, and on the dependence of the pressure on the number of particles. We investigate the consequence of increasing the number of particles and verify that the giant number fluctuations result in an increase of the extractable work, in contrast to an ideal gas many-body Szil\'ard engine. Finally, we experimentally extract work from the Szil\'ard engine by lifting a mass against gravity. We compare the measured work to the expected work within the constraints of the protocol, and to the work extractable in an optimal (quasistatic) process.

Particles in our system are self-propelled, elongated bristle-robots (Hexbug, Nano). Each robot has $12$ flexible legs on which it jumps due to the rotation of an internal, battery-powered motor \cite{giomi2013swarming}. Bristle-robots (bbots) exhibit typical active matter behaviors such as clustering \cite{giomi2013swarming, deblais2018boundaries, sanchez2018self}. The motion of a single bbot can be described as directed random motion with a persistence length of more than $1$~m \cite{dauchot2019dynamics}, which is larger than the radius of our system, $R\simeq0.2$~m. Bbots exhibit a slightly chiral motion that did not significantly affect the resulting pressure and particle distribution.  We place the bbots in a half-circular arena with rounded edges to prevent particles from getting stuck in corners. The arena has a detachable partition that can be inserted at will and rotates freely on an axle. We record the experiments with a web camera (Logitech, Brio 4K).

{\it Pressure as a function of density} - The relation between pressure and density is determined by measuring the average steady-state volume occupied by the bbots under a specified external pressure. A constant force is applied by attaching a mass $m$ to the partition. The bots are positioned on one side of the partition and exert pressure to counteract the gravitational force (see SI Movie 2 \cite{comment0}). The mechanical quasi-two-dimensional pressure that the bbots apply on the partition is given by $mg/L$, where $g$ is the gravitational acceleration, and $L$ is the length of the partition. The density, at this pressure, is given by the number of bbots in the chamber divided by the average volume they occupy.

\begin{figure}
    \centering
    \includegraphics{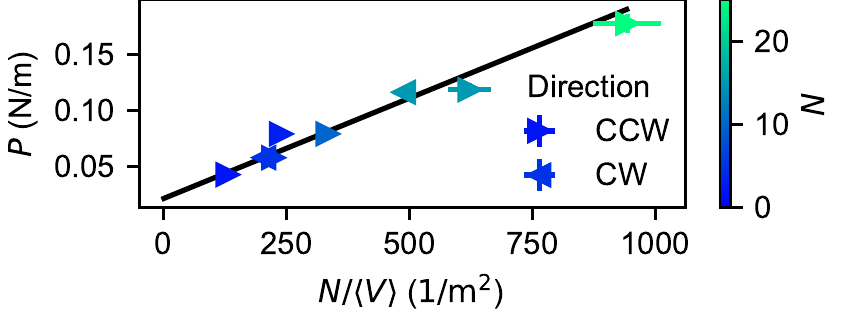}
    \caption{Applied pressure as a function of density. A linear fit, $P=\alpha N/V + P_0$, is shown in a black line. Different colors indicate different particle numbers $N$. To rule out a possible effect of the chiral motion of bbots, measurements were made with particles at either side of the partition (triangular symbols indicate the direction of applied force on the partition). }
    \label{fig:pressure_density}
\end{figure}

In Fig.~\ref{fig:pressure_density}, we plot the applied pressure as a function of the average measured density $N/V$, demonstrating that pressure is a monotonous function that grows with density. We fit the pressure to an ideal-gas-like equation of state
$P = \alpha N/V + P_0$,
with a fitting parameter $\alpha=1.78\pm0.21\times10^{-4}$~J (Fig.~\ref{fig:pressure_density}, solid line). The nonzero intercept ($P_0 = 2.18 \pm 0.52 \times 10^{-2}$~N/m) is attributed to a combination of static friction and the finite size of the system. At high loads, the finite system size restricts the partition and volume fluctuations, truncating its range. This leads to an underestimation of the mechanical equilibrium volume and an overestimation of the pressure in a non-trivial manner (see SI section 2 \cite{comment0}). It should be noted that the intercept is canceled out in subsequent calculations that are based on the difference of pressures from both sides.

\begin{figure}
    \centering
    \includegraphics{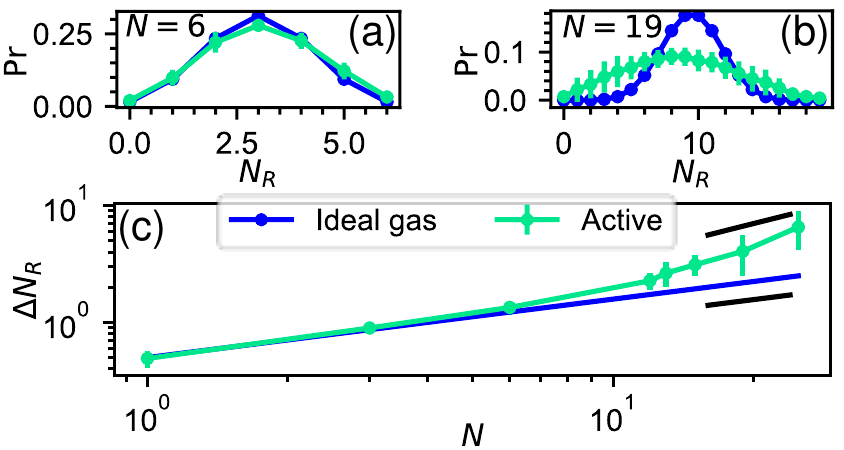}
    \caption{Experimental measurement of number fluctuations. \textbf{a-b}) Histograms of the number of particles in the right half of the system $N_R$ for $N=6,19$ particles (green line) respectively,  compared to the expected distribution for an ideal gas (blue line).
    (\textbf{c}) The standard deviation of the distribution of the number of particles in the right half of the system $\Delta N_R$ as a function of the mean amount of particles in it $\langle N_R\rangle$. The black lines are the power laws for a system in equilibrium ($\Delta N_R \sim {N}^{1/2}$, bottom) and for a system with giant number fluctuations ($\Delta N_R \sim N$, top).
    We observe a transition between these regimes when the number of particles is increased.}
    \label{fig:num_fluctuations}
\end{figure}

{\it
 Number fluctuations} - We use image analysis \cite{bradski2000opencv} to measure the distribution of the number of particles in each half, e.g. $\Pr (N_R)$. For small $N$ the distribution is essentially that of ideal thermal particles (Fig. \ref{fig:num_fluctuations}a). In contrast, much wider distributions are observed for larger $N_R$ (Fig. \ref{fig:num_fluctuations}b).
 In Fig.~\ref{fig:num_fluctuations}c, we show the standard deviation $\Delta N_R$ of $N_R$ for different values of $N$. The results exhibit a transition from $\Delta N_R \propto \sqrt{N}$ towards the maximally possible exponent, $\Delta N_R \propto N$. Such high number fluctuations were found for other rod-like active materials \cite{Ramaswamy2003, toner2005hydrodynamics,chate2008collective, ginelli2010large}.
 
\begin{figure}
    \centering
    \includegraphics[width=\linewidth]{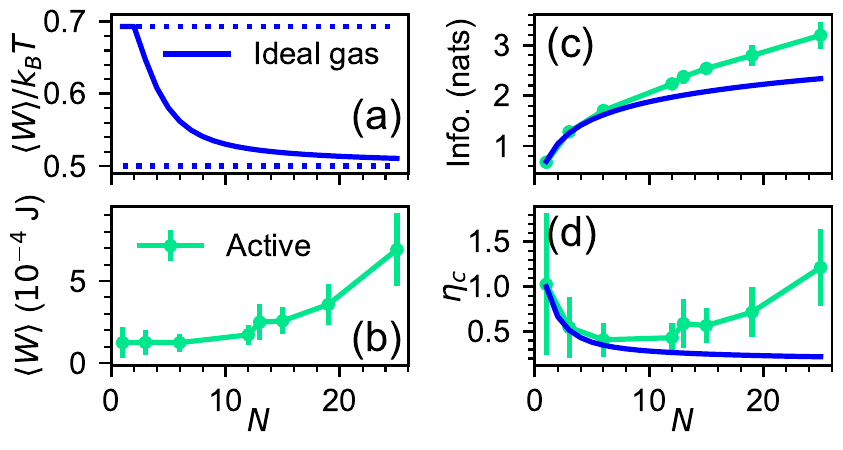}
    \caption{Upper bound on mean work per cycle as a function of the number of particles for the many-body Szil\'ard's engines: (\textbf{a}) passive (solid blue line) and (\textbf{b}) active (green points). Values for the ideal gas are normalized by $k_BT$ whereas, for the active gas, work is in units of $10^{-4}$ Joules. (\textbf{c}) Information per measurement as a function of $N$. (\textbf{d}) The efficiency of converting information to work, for quasistatic work extraction, as a function of $N$.}
    \label{fig:average_work}
\end{figure}

{\it Maximal extractable work} - Keeping in mind Szil\'ard's original view, we consider a case where work is extracted by connecting a mass that is lifted against gravity.
The work $W=\int F dh$ clearly depends on both the time-dependent height $h$ of the mass and the force $F$ applied at every instant. In equilibrium, optimal work extraction is achieved by following a quasi-static process in which the force balances the pressure difference at any time. In this case, all the free energy difference is converted into work. This quasi-static work is given by
\begin{equation}
    W(k) = \int_{\frac{1}{2}V_{tot}}^{\frac{N/2+k}{N}V_{tot}} P_R dV_R + \int_{\frac{1}{2}V_{tot}}^{\frac{N/2-k}{N}V_{tot}} P_L dV_L,
\end{equation}
where $k=N_R - N/2$ and $P_{L,R}$ are the pressure of the left/right-hand side of the system. For an ideal gas,
\begin{equation}\label{eq:ideal_gas_work}
    W(k) = N k_B T \left( 
    \frac{1}{2} \ln\left(1-4\left(\frac{k}{N}\right)^2\right) + \frac{k}{N} \ln\frac{1+2\frac{k}{N}}{1-2\frac{k}{N}}
    \right),
\end{equation}
where $T$ is the temperature and $k_B$ is Boltzmann's constant.
The average extracted work given that particles are non-interacting and their spatial distributions are uniform, is 
$\langle W \rangle = \sum_{k = -N/2}^{N/2} W(k) \Pr(k)$, with $\Pr(k) = \left(\frac{1}{2}\right)^N \binom{N}{\frac{N}{2}+k}$. For an engine operating on an ideal gas and  $N\gg1$, the mean work can approximated by noting that $\Pr(k)$ is a fast decaying function as $|k/N|$ grows. As a result one can expand $W(k)$ \cite{kim2011information},
\begin{align}
    &W(k) = N k_B T \left(2\left(\frac{k}{N}\right)^2 + O \left(\left(\frac{k}{N}\right)^4\right)\right), \\
    &\lim_{N\to\infty}\langle W \rangle = \frac{2 k_B T}{N} \langle k^2 \rangle = \frac{1}{2} k_B T, \label{eq:work_asympt}
\end{align}
and higher-order terms vanish. 
Therefore, the mean extractable work depends on number fluctuations ($\langle k^2\rangle = (\Delta N_R)^2$), and for an ideal-gas-based Szil\'ard engine it transitions from  $k_B T \ln 2$ to $k_B T / 2$ as $N$ increases (Fig.~\ref{fig:average_work}a).

We calculate $\langle W \rangle$  for the active particle system following the same analysis, substituting $\alpha$ (Fig.~\ref{fig:pressure_density}) for $k_B T$ and the experimentally measured distribution $\Pr(k)$ in place of the binomial distribution (Fig.~\ref{fig:average_work}b). We find that the calculated $\langle W \rangle$ increases with $N$. This is a clear qualitative difference between information engines operating on active and thermal particles.

{\it Information and efficiency} - The protocol used to extract work uses information about the number of particles in each half to determine the weight that can be lifted. The information measure that fits this protocol is the Shannon information of the distribution $\Pr (N_R)$, $I=-\sum_{N_R=0}^{N} \Pr(N_R) \ln  \Pr(N_R)$. For an engine with thermal particles, the efficiency of converting information to work is $\eta_c = \langle W \rangle / k_B T I$. For comparison, we use a figure of merit $\eta_c = \langle W \rangle / \alpha I$ also for our active information engine.

The information and $\eta_c$ for both cases are shown in Fig. \ref{fig:average_work}c,d. The wider number distribution of the active particles is reflected in larger $I$. At the same time, the calculated $\eta_c$ for the active particles is larger than for their thermal counterparts, especially for large $N$. In fact, $\eta_c>1$ is found for the active system with $N=25$. It should be stressed that in the active system, there is an additional source of energy that is not taken into account in the definition of $\eta_c$, which is therefore not the true thermodynamic efficiency of the setup.

{\it Experimental work extraction} - In theoretical analyses of information engines, work is commonly assumed to be extracted in an optimal way and more attention is directed to finding the optimal measurement protocol \cite{horowitz2011designing, park2016optimal, song2021optimal}.
Extraction of the full amount of energy from the post-measurement state is an experimental challenge, requiring
the ability to control the system quasi-statically \cite{toyabe2010experimental, berut2012experimental, ribezzi2019large}. 
In many cases, a slow quasi-static work extraction process is undesired and impractical, as is in our experimental setup which has long relaxation times and is controlled manually. Taking these considerations into account, we demonstrate that measurable work can be extracted from our system by testing a finite-time approximation of the optimal protocol \cite{andresen1977thermodynamics}. 

Work extraction is demonstrated using the following protocol. The partition is placed in the center of the arena, with different numbers of bbots on its sides, thereby matching a particular measurement of $N_L, N_R$. A string with three weights is then attached to the partition. The weights are chosen so their sum partially balances the expected pressure imbalance. The partition is then allowed to move freely so that after a while it fluctuates around the volume ratio where the forces on it are balanced. At this point, the lower weight ($m_1$) is removed by burning the string connecting it to the other weights (see SI Movie 3 \cite{comment0}). The lower net weight on the string means that the partition is free to move and fluctuate around a new balancing point while raising the remaining weights. The process is repeated. At the end of the process, the i-th weight has been raised by $\Delta h_i$. The work extracted in the process is given by $W= g \sum_i m_i \Delta h_i$. The partition exhibits large fluctuations, but repetitions of the process allow for approximation of the ensemble average of the work. Clearly, the protocol can be refined by adding more steps with smaller masses, where quasi-static control will be achieved at the limit of infinitely many infinitesimally small masses. 

\begin{figure}
    \centering
    \includegraphics[width=\linewidth]{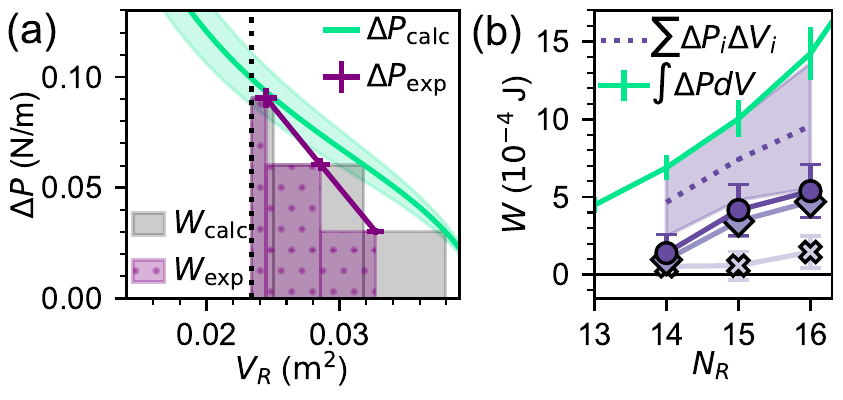}
    \caption{Experimental and predicted work extraction for the three-step expansion protocol compared to the quasi-static limits. (\textbf{a}) Pressure difference applied on the partition as a function of the right half's volume $V_R$, at $N_R/N=15/17$. The calculated pressure difference $\Delta P$ on the partition, based on data from Fig.~\ref{fig:pressure_density}, is shown in a solid green line. The experimental results are shown in purple points.
    The initial volume is $V_R=V/2$, shown in a dotted vertical line. The work extracted at each step is $W=\Delta P \Delta V$ (area of gray and purple rectangles for calculated and experimental results, respectively).
    (\textbf{b}) Work extracted from the system for experiments with different values of $N_R/N$ for $N=17$ after one (crosses), two (diamonds) and three (circles) stages of the protocol. For $N_R=15$, values correspond to the areas of purple rectangles in (\textbf{a}). Results are compared to the expected work from the finite time protocol with three control steps (dashed purple line), and the optimal control protocol (solid green line).}
    \label{fig:protocols}
\end{figure}

In Fig.~\ref{fig:protocols}b we plot the upper bound of the extractable work $W(k)$ (solid green line), calculated for a quasi-static process, the expected work (see SI section 2 \cite{comment0}) for the three-stage protocol (dashed purple lines), and the measured work (markers) for experiments with $N=17$ and $N_R=14,15,16$.
The measured work exhibits the same trends as its expected counterpart and also falls just within experimental uncertainty. However, it also is systematically smaller than expected. This difference is attributed to a mechanical constraint that restricts the maximally attainable volume (see SI section 2 \cite{comment0}). Nevertheless, the results show that more work can be extracted when the imbalance between the number of particles in the two halves is larger, and as more control steps are added. This verifies that pressure measurements are reliable and can be used to calculate the amount of extracted work for a given work extraction protocol, hence demonstrating that work of order $10^{-4}$~J can be extracted from the engine, as predicted in Fig.~\ref{fig:average_work}.

In summary, we report the first experimental study of a macroscopic information engine operating on an active system. Our implementation was based on a many-body  Szil\'ard's engine configuration consisting of self-propelled bbots. We analyzed the operation of the engine in comparison to one based on an ideal gas. The differences between the two reveal themselves in the dependence of the work per cycle on the number of particles. Specifically, the extractable work increases with particle number in our active engine, while decreasing in an ideal gas engine. This fundamental difference is attributed to the giant number fluctuations exhibited by the active system. This is a general result and should hold for any system with giant number fluctuations and monotonous dependence of pressure on density. While previous studies of work extraction from active systems were based on the interaction between active matter and boundaries \cite{hiratsuka2006microrotary, angelani2009self, di2010bacterial, angelani2011active, lagoin2022human}, here we use a different attribute of active matter: their tendency to aggregate.

In addition, we demonstrated how different protocols affect the extracted work. It would be interesting to explore different protocols to maximize work extraction at finite operation times, especially in light of the literature on optimal control of active matter \cite{gupta2023efficient, norton2020optimal, shankar2022optimal, yang2018optimal}.  Clearly, more work could be extracted from our system using more elaborated feedback and control schemes that employ information about the dynamics of the partition during the expansion process.

Fluctuations in our experiment are athermal and can occur at any scale, without the amplification required in \cite{freitas2022maxwell}. This allowed us to extract work with an order of magnitude $\alpha\simeq10^{-4}$~J per cycle, where $\alpha$ is the scaling factor between pressure and density for the bbots (Fig.~\ref{fig:pressure_density}). In comparison, state-of-the-art microscopic information engines can only extract up to about $k_BT\simeq10^{-21}$~J per cycle \cite{ribezzi2019large, saha2021maximizing}, or $10 k_B T\simeq 10^{-20}$~J when a passive particle is subjected to nonequilibrium fluctuations \cite{saha2023information, paneru2022colossal}. Notably, the energy scale $\alpha$ in our experiment is also three orders of magnitude larger than that measured for a macroscopic Maxwell's demon operating on a granular gas \cite{lagoin2022human}.

To compare our results to microscopic information engines on equal footings, we also evaluate the normalized values of work and efficiency. The ideal work per cycle (Fig.~\ref{fig:average_work}) is $\langle W \rangle/\alpha\simeq3.88$ at $N=25$, which is comparable to $\langle W \rangle/k_B T \simeq 1$ values in microscopic engines \cite{ribezzi2019large, paneru2018lossless, saha2021maximizing}. The corresponding maximum normalized information-to-work conversion efficiency is $\eta_c=\langle W \rangle/\alpha I=1.21\pm 0.42$, which is of course a result of the energy required to maintain activity not being taken into account. A conversion efficiency $\eta_c>1$ highlights the non-equilibrium dynamics the active particles. This is in line with the results of \cite{saha2023information} but in stark contrast to the situation in \cite{paneru2022colossal}, where the non-equilibrium fluctuations are analogous to thermal fluctuations at a high effective temperature.

Finally, we note that when considering the energy required to sustain the active system in a steady state, the full thermodynamic efficiency of our active information engine is naturally very low. This is a result of only a fraction of the total energy being used to push the partition. Evidently, directly connecting the batteries to an electric engine would yield a significantly higher power output than can be extracted from the movement of the partition. However, understanding the various mechanisms by which energy can be extracted from active matter is fundamentally interesting. It will also assist in the design of microscopic engines operating in naturally occurring active systems, such as bacteria swarms.

\begin{acknowledgements}

We thank Isaac Yekotiel and the Chemistry Machine Shop of Tel Aviv University for the design and construction of the experimental system.
OC and YR acknowledge support from the European Research Council (ERC) under the
European Union’s Horizon 2020 research and innovation programme (Grant agreement No. 101002392). AS and YR acknowledge support from the Israel Science Foundation (Grant No. 385/21). SR is grateful for support from the Israel Science Foundation (Grant No. 1929/21). RG acknowledges support from the Ratner Center for single molecule sciences.
\end{acknowledgements}


\begin{thebibliography}{63}%
\makeatletter
\providecommand \@ifxundefined [1]{%
 \@ifx{#1\undefined}
}%
\providecommand \@ifnum [1]{%
 \ifnum #1\expandafter \@firstoftwo
 \else \expandafter \@secondoftwo
 \fi
}%
\providecommand \@ifx [1]{%
 \ifx #1\expandafter \@firstoftwo
 \else \expandafter \@secondoftwo
 \fi
}%
\providecommand \natexlab [1]{#1}%
\providecommand \enquote  [1]{``#1''}%
\providecommand \bibnamefont  [1]{#1}%
\providecommand \bibfnamefont [1]{#1}%
\providecommand \citenamefont [1]{#1}%
\providecommand \href@noop [0]{\@secondoftwo}%
\providecommand \href [0]{\begingroup \@sanitize@url \@href}%
\providecommand \@href[1]{\@@startlink{#1}\@@href}%
\providecommand \@@href[1]{\endgroup#1\@@endlink}%
\providecommand \@sanitize@url [0]{\catcode `\\12\catcode `\$12\catcode
  `\&12\catcode `\#12\catcode `\^12\catcode `\_12\catcode `\%12\relax}%
\providecommand \@@startlink[1]{}%
\providecommand \@@endlink[0]{}%
\providecommand \url  [0]{\begingroup\@sanitize@url \@url }%
\providecommand \@url [1]{\endgroup\@href {#1}{\urlprefix }}%
\providecommand \urlprefix  [0]{URL }%
\providecommand \Eprint [0]{\href }%
\providecommand \doibase [0]{https://doi.org/}%
\providecommand \selectlanguage [0]{\@gobble}%
\providecommand \bibinfo  [0]{\@secondoftwo}%
\providecommand \bibfield  [0]{\@secondoftwo}%
\providecommand \translation [1]{[#1]}%
\providecommand \BibitemOpen [0]{}%
\providecommand \bibitemStop [0]{}%
\providecommand \bibitemNoStop [0]{.\EOS\space}%
\providecommand \EOS [0]{\spacefactor3000\relax}%
\providecommand \BibitemShut  [1]{\csname bibitem#1\endcsname}%
\let\auto@bib@innerbib\@empty
\bibitem [{\citenamefont {Parrondo}\ \emph {et~al.}(2015)\citenamefont
  {Parrondo}, \citenamefont {Horowitz},\ and\ \citenamefont
  {Sagawa}}]{parrondo2015thermodynamics}%
  \BibitemOpen
  \bibfield  {author} {\bibinfo {author} {\bibfnamefont {J.~M.}\ \bibnamefont
  {Parrondo}}, \bibinfo {author} {\bibfnamefont {J.~M.}\ \bibnamefont
  {Horowitz}},\ and\ \bibinfo {author} {\bibfnamefont {T.}~\bibnamefont
  {Sagawa}},\ }\bibfield  {title} {\bibinfo {title} {Thermodynamics of
  information},\ }\href@noop {} {\bibfield  {journal} {\bibinfo  {journal}
  {Nat. Phys.}\ }\textbf {\bibinfo {volume} {11}},\ \bibinfo {pages} {131}
  (\bibinfo {year} {2015})}\BibitemShut {NoStop}%
\bibitem [{\citenamefont {Landauer}(1961)}]{landauer1961irreversibility}%
  \BibitemOpen
  \bibfield  {author} {\bibinfo {author} {\bibfnamefont {R.}~\bibnamefont
  {Landauer}},\ }\bibfield  {title} {\bibinfo {title} {Irreversibility and heat
  generation in the computing process},\ }\href@noop {} {\bibfield  {journal}
  {\bibinfo  {journal} {IBM J. Res. Dev.}\ }\textbf {\bibinfo {volume} {5}},\
  \bibinfo {pages} {183} (\bibinfo {year} {1961})}\BibitemShut {NoStop}%
\bibitem [{\citenamefont {Bennett}(1973)}]{bennett1973logical}%
  \BibitemOpen
  \bibfield  {author} {\bibinfo {author} {\bibfnamefont {C.~H.}\ \bibnamefont
  {Bennett}},\ }\bibfield  {title} {\bibinfo {title} {Logical reversibility of
  computation},\ }\href@noop {} {\bibfield  {journal} {\bibinfo  {journal} {IBM
  J. Res. Dev.}\ }\textbf {\bibinfo {volume} {17}},\ \bibinfo {pages} {525}
  (\bibinfo {year} {1973})}\BibitemShut {NoStop}%
\bibitem [{\citenamefont {Sagawa}\ and\ \citenamefont
  {Ueda}(2008)}]{sagawa2008second}%
  \BibitemOpen
  \bibfield  {author} {\bibinfo {author} {\bibfnamefont {T.}~\bibnamefont
  {Sagawa}}\ and\ \bibinfo {author} {\bibfnamefont {M.}~\bibnamefont {Ueda}},\
  }\bibfield  {title} {\bibinfo {title} {Second law of thermodynamics with
  discrete quantum feedback control},\ }\href@noop {} {\bibfield  {journal}
  {\bibinfo  {journal} {Phys. Rev. Lett.}\ }\textbf {\bibinfo {volume} {100}},\
  \bibinfo {pages} {080403} (\bibinfo {year} {2008})}\BibitemShut {NoStop}%
\bibitem [{\citenamefont {Cao}\ and\ \citenamefont
  {Feito}(2009)}]{cao2009thermodynamics}%
  \BibitemOpen
  \bibfield  {author} {\bibinfo {author} {\bibfnamefont {F.~J.}\ \bibnamefont
  {Cao}}\ and\ \bibinfo {author} {\bibfnamefont {M.}~\bibnamefont {Feito}},\
  }\bibfield  {title} {\bibinfo {title} {Thermodynamics of feedback controlled
  systems},\ }\href@noop {} {\bibfield  {journal} {\bibinfo  {journal} {Phys.
  Rev. E}\ }\textbf {\bibinfo {volume} {79}},\ \bibinfo {pages} {041118}
  (\bibinfo {year} {2009})}\BibitemShut {NoStop}%
\bibitem [{\citenamefont {Horowitz}\ and\ \citenamefont
  {Vaikuntanathan}(2010)}]{horowitz2010nonequilibrium}%
  \BibitemOpen
  \bibfield  {author} {\bibinfo {author} {\bibfnamefont {J.~M.}\ \bibnamefont
  {Horowitz}}\ and\ \bibinfo {author} {\bibfnamefont {S.}~\bibnamefont
  {Vaikuntanathan}},\ }\bibfield  {title} {\bibinfo {title} {Nonequilibrium
  detailed fluctuation theorem for repeated discrete feedback},\ }\href@noop {}
  {\bibfield  {journal} {\bibinfo  {journal} {Phys. Rev. E}\ }\textbf {\bibinfo
  {volume} {82}},\ \bibinfo {pages} {061120} (\bibinfo {year}
  {2010})}\BibitemShut {NoStop}%
\bibitem [{\citenamefont {Sagawa}\ and\ \citenamefont
  {Ueda}(2010)}]{sagawa2010generalized}%
  \BibitemOpen
  \bibfield  {author} {\bibinfo {author} {\bibfnamefont {T.}~\bibnamefont
  {Sagawa}}\ and\ \bibinfo {author} {\bibfnamefont {M.}~\bibnamefont {Ueda}},\
  }\bibfield  {title} {\bibinfo {title} {Generalized jarzynski equality under
  nonequilibrium feedback control},\ }\href@noop {} {\bibfield  {journal}
  {\bibinfo  {journal} {Phys. Rev. Lett.}\ }\textbf {\bibinfo {volume} {104}},\
  \bibinfo {pages} {090602} (\bibinfo {year} {2010})}\BibitemShut {NoStop}%
\bibitem [{\citenamefont {Ponmurugan}(2010)}]{ponmurugan2010generalized}%
  \BibitemOpen
  \bibfield  {author} {\bibinfo {author} {\bibfnamefont {M.}~\bibnamefont
  {Ponmurugan}},\ }\bibfield  {title} {\bibinfo {title} {Generalized detailed
  fluctuation theorem under nonequilibrium feedback control},\ }\href@noop {}
  {\bibfield  {journal} {\bibinfo  {journal} {Phys. Rev. E}\ }\textbf {\bibinfo
  {volume} {82}},\ \bibinfo {pages} {031129} (\bibinfo {year}
  {2010})}\BibitemShut {NoStop}%
\bibitem [{\citenamefont {Fujitani}\ and\ \citenamefont
  {Suzuki}(2010)}]{fujitani2010jarzynski}%
  \BibitemOpen
  \bibfield  {author} {\bibinfo {author} {\bibfnamefont {Y.}~\bibnamefont
  {Fujitani}}\ and\ \bibinfo {author} {\bibfnamefont {H.}~\bibnamefont
  {Suzuki}},\ }\bibfield  {title} {\bibinfo {title} {Jarzynski equality
  modified in the linear feedback system},\ }\href@noop {} {\bibfield
  {journal} {\bibinfo  {journal} {J. Phys. Soc. Jpn.}\ }\textbf {\bibinfo
  {volume} {79}},\ \bibinfo {pages} {104003} (\bibinfo {year}
  {2010})}\BibitemShut {NoStop}%
\bibitem [{\citenamefont {Kundu}(2012)}]{kundu2012nonequilibrium}%
  \BibitemOpen
  \bibfield  {author} {\bibinfo {author} {\bibfnamefont {A.}~\bibnamefont
  {Kundu}},\ }\bibfield  {title} {\bibinfo {title} {Nonequilibrium fluctuation
  theorem for systems under discrete and continuous feedback control},\
  }\href@noop {} {\bibfield  {journal} {\bibinfo  {journal} {Phys. Rev. E}\
  }\textbf {\bibinfo {volume} {86}},\ \bibinfo {pages} {021107} (\bibinfo
  {year} {2012})}\BibitemShut {NoStop}%
\bibitem [{\citenamefont {Maxwell}(1871)}]{maxwell1871theory}%
  \BibitemOpen
  \bibfield  {author} {\bibinfo {author} {\bibfnamefont {J.~C.}\ \bibnamefont
  {Maxwell}},\ }\bibfield  {title} {\bibinfo {title} {Theory of heat},\
  }\href@noop {} {\bibfield  {journal} {\bibinfo  {journal} {London, UK:
  Longmans}\ } (\bibinfo {year} {1871})}\BibitemShut {NoStop}%
\bibitem [{\citenamefont {Admon}\ \emph {et~al.}(2018)\citenamefont {Admon},
  \citenamefont {Rahav},\ and\ \citenamefont
  {Roichman}}]{admon2018experimental}%
  \BibitemOpen
  \bibfield  {author} {\bibinfo {author} {\bibfnamefont {T.}~\bibnamefont
  {Admon}}, \bibinfo {author} {\bibfnamefont {S.}~\bibnamefont {Rahav}},\ and\
  \bibinfo {author} {\bibfnamefont {Y.}~\bibnamefont {Roichman}},\ }\bibfield
  {title} {\bibinfo {title} {Experimental realization of an information machine
  with tunable temporal correlations},\ }\href@noop {} {\bibfield  {journal}
  {\bibinfo  {journal} {Phys. Rev. Lett.}\ }\textbf {\bibinfo {volume} {121}},\
  \bibinfo {pages} {180601} (\bibinfo {year} {2018})}\BibitemShut {NoStop}%
\bibitem [{\citenamefont {Jun}\ \emph {et~al.}(2014)\citenamefont {Jun},
  \citenamefont {Gavrilov},\ and\ \citenamefont {Bechhoefer}}]{jun2014high}%
  \BibitemOpen
  \bibfield  {author} {\bibinfo {author} {\bibfnamefont {Y.}~\bibnamefont
  {Jun}}, \bibinfo {author} {\bibfnamefont {M.}~\bibnamefont {Gavrilov}},\ and\
  \bibinfo {author} {\bibfnamefont {J.}~\bibnamefont {Bechhoefer}},\ }\bibfield
   {title} {\bibinfo {title} {High-precision test of landauer’s principle in
  a feedback trap},\ }\href@noop {} {\bibfield  {journal} {\bibinfo  {journal}
  {Phys. Rev. Lett.}\ }\textbf {\bibinfo {volume} {113}},\ \bibinfo {pages}
  {190601} (\bibinfo {year} {2014})}\BibitemShut {NoStop}%
\bibitem [{\citenamefont {B{\'e}rut}\ \emph {et~al.}(2012)\citenamefont
  {B{\'e}rut}, \citenamefont {Arakelyan}, \citenamefont {Petrosyan},
  \citenamefont {Ciliberto}, \citenamefont {Dillenschneider},\ and\
  \citenamefont {Lutz}}]{berut2012experimental}%
  \BibitemOpen
  \bibfield  {author} {\bibinfo {author} {\bibfnamefont {A.}~\bibnamefont
  {B{\'e}rut}}, \bibinfo {author} {\bibfnamefont {A.}~\bibnamefont
  {Arakelyan}}, \bibinfo {author} {\bibfnamefont {A.}~\bibnamefont
  {Petrosyan}}, \bibinfo {author} {\bibfnamefont {S.}~\bibnamefont
  {Ciliberto}}, \bibinfo {author} {\bibfnamefont {R.}~\bibnamefont
  {Dillenschneider}},\ and\ \bibinfo {author} {\bibfnamefont {E.}~\bibnamefont
  {Lutz}},\ }\bibfield  {title} {\bibinfo {title} {Experimental verification of
  landauer’s principle linking information and thermodynamics},\ }\href@noop
  {} {\bibfield  {journal} {\bibinfo  {journal} {Nature}\ }\textbf {\bibinfo
  {volume} {483}},\ \bibinfo {pages} {187} (\bibinfo {year}
  {2012})}\BibitemShut {NoStop}%
\bibitem [{\citenamefont {Toyabe}\ \emph {et~al.}(2010)\citenamefont {Toyabe},
  \citenamefont {Sagawa}, \citenamefont {Ueda}, \citenamefont {Muneyuki},\ and\
  \citenamefont {Sano}}]{toyabe2010experimental}%
  \BibitemOpen
  \bibfield  {author} {\bibinfo {author} {\bibfnamefont {S.}~\bibnamefont
  {Toyabe}}, \bibinfo {author} {\bibfnamefont {T.}~\bibnamefont {Sagawa}},
  \bibinfo {author} {\bibfnamefont {M.}~\bibnamefont {Ueda}}, \bibinfo {author}
  {\bibfnamefont {E.}~\bibnamefont {Muneyuki}},\ and\ \bibinfo {author}
  {\bibfnamefont {M.}~\bibnamefont {Sano}},\ }\bibfield  {title} {\bibinfo
  {title} {Experimental demonstration of information-to-energy conversion and
  validation of the generalized jarzynski equality},\ }\href@noop {} {\bibfield
   {journal} {\bibinfo  {journal} {Nat. Phys.}\ }\textbf {\bibinfo {volume}
  {6}},\ \bibinfo {pages} {988} (\bibinfo {year} {2010})}\BibitemShut {NoStop}%
\bibitem [{\citenamefont {Ribezzi-Crivellari}\ and\ \citenamefont
  {Ritort}(2019)}]{ribezzi2019large}%
  \BibitemOpen
  \bibfield  {author} {\bibinfo {author} {\bibfnamefont {M.}~\bibnamefont
  {Ribezzi-Crivellari}}\ and\ \bibinfo {author} {\bibfnamefont
  {F.}~\bibnamefont {Ritort}},\ }\bibfield  {title} {\bibinfo {title} {Large
  work extraction and the landauer limit in a continuous maxwell demon},\
  }\href@noop {} {\bibfield  {journal} {\bibinfo  {journal} {Nat. Phys.}\
  }\textbf {\bibinfo {volume} {15}},\ \bibinfo {pages} {660} (\bibinfo {year}
  {2019})}\BibitemShut {NoStop}%
\bibitem [{\citenamefont {Paneru}\ \emph {et~al.}(2018)\citenamefont {Paneru},
  \citenamefont {Lee}, \citenamefont {Tlusty},\ and\ \citenamefont
  {Pak}}]{paneru2018lossless}%
  \BibitemOpen
  \bibfield  {author} {\bibinfo {author} {\bibfnamefont {G.}~\bibnamefont
  {Paneru}}, \bibinfo {author} {\bibfnamefont {D.~Y.}\ \bibnamefont {Lee}},
  \bibinfo {author} {\bibfnamefont {T.}~\bibnamefont {Tlusty}},\ and\ \bibinfo
  {author} {\bibfnamefont {H.~K.}\ \bibnamefont {Pak}},\ }\bibfield  {title}
  {\bibinfo {title} {Lossless brownian information engine},\ }\href@noop {}
  {\bibfield  {journal} {\bibinfo  {journal} {Phys. Rev. Lett.}\ }\textbf
  {\bibinfo {volume} {120}},\ \bibinfo {pages} {020601} (\bibinfo {year}
  {2018})}\BibitemShut {NoStop}%
\bibitem [{\citenamefont {Saha}\ \emph {et~al.}(2021)\citenamefont {Saha},
  \citenamefont {Lucero}, \citenamefont {Ehrich}, \citenamefont {Sivak},\ and\
  \citenamefont {Bechhoefer}}]{saha2021maximizing}%
  \BibitemOpen
  \bibfield  {author} {\bibinfo {author} {\bibfnamefont {T.~K.}\ \bibnamefont
  {Saha}}, \bibinfo {author} {\bibfnamefont {J.~N.}\ \bibnamefont {Lucero}},
  \bibinfo {author} {\bibfnamefont {J.}~\bibnamefont {Ehrich}}, \bibinfo
  {author} {\bibfnamefont {D.~A.}\ \bibnamefont {Sivak}},\ and\ \bibinfo
  {author} {\bibfnamefont {J.}~\bibnamefont {Bechhoefer}},\ }\bibfield  {title}
  {\bibinfo {title} {Maximizing power and velocity of an information engine},\
  }\href@noop {} {\bibfield  {journal} {\bibinfo  {journal} {Proc. Nat. Acad.
  Sci.}\ }\textbf {\bibinfo {volume} {118}},\ \bibinfo {pages} {e2023356118}
  (\bibinfo {year} {2021})}\BibitemShut {NoStop}%
\bibitem [{\citenamefont {Paneru}\ and\ \citenamefont
  {Kyu~Pak}(2020)}]{paneru2020colloidal}%
  \BibitemOpen
  \bibfield  {author} {\bibinfo {author} {\bibfnamefont {G.}~\bibnamefont
  {Paneru}}\ and\ \bibinfo {author} {\bibfnamefont {H.}~\bibnamefont
  {Kyu~Pak}},\ }\bibfield  {title} {\bibinfo {title} {Colloidal engines for
  innovative tests of information thermodynamics},\ }\href@noop {} {\bibfield
  {journal} {\bibinfo  {journal} {Adv. Phys.: X}\ }\textbf {\bibinfo {volume}
  {5}},\ \bibinfo {pages} {1823880} (\bibinfo {year} {2020})}\BibitemShut
  {NoStop}%
\bibitem [{\citenamefont {Koski}\ \emph
  {et~al.}(2014{\natexlab{a}})\citenamefont {Koski}, \citenamefont {Maisi},
  \citenamefont {Sagawa},\ and\ \citenamefont {Pekola}}]{koski2014mutual}%
  \BibitemOpen
  \bibfield  {author} {\bibinfo {author} {\bibfnamefont {J.~V.}\ \bibnamefont
  {Koski}}, \bibinfo {author} {\bibfnamefont {V.~F.}\ \bibnamefont {Maisi}},
  \bibinfo {author} {\bibfnamefont {T.}~\bibnamefont {Sagawa}},\ and\ \bibinfo
  {author} {\bibfnamefont {J.~P.}\ \bibnamefont {Pekola}},\ }\bibfield  {title}
  {\bibinfo {title} {Experimental observation of the role of mutual information
  in the nonequilibrium dynamics of a maxwell demon},\ }\href@noop {}
  {\bibfield  {journal} {\bibinfo  {journal} {Phys. Rev. Lett.}\ }\textbf
  {\bibinfo {volume} {113}},\ \bibinfo {pages} {030601} (\bibinfo {year}
  {2014}{\natexlab{a}})}\BibitemShut {NoStop}%
\bibitem [{\citenamefont {Koski}\ \emph
  {et~al.}(2014{\natexlab{b}})\citenamefont {Koski}, \citenamefont {Maisi},
  \citenamefont {Pekola},\ and\ \citenamefont
  {Averin}}]{koski2014experimental}%
  \BibitemOpen
  \bibfield  {author} {\bibinfo {author} {\bibfnamefont {J.~V.}\ \bibnamefont
  {Koski}}, \bibinfo {author} {\bibfnamefont {V.~F.}\ \bibnamefont {Maisi}},
  \bibinfo {author} {\bibfnamefont {J.~P.}\ \bibnamefont {Pekola}},\ and\
  \bibinfo {author} {\bibfnamefont {D.~V.}\ \bibnamefont {Averin}},\ }\bibfield
   {title} {\bibinfo {title} {Experimental realization of a szilard engine with
  a single electron},\ }\href@noop {} {\bibfield  {journal} {\bibinfo
  {journal} {Proc. Nat. Acad. Sci.}\ }\textbf {\bibinfo {volume} {111}},\
  \bibinfo {pages} {13786} (\bibinfo {year} {2014}{\natexlab{b}})}\BibitemShut
  {NoStop}%
\bibitem [{\citenamefont {Masuyama}\ \emph {et~al.}(2018)\citenamefont
  {Masuyama}, \citenamefont {Funo}, \citenamefont {Murashita}, \citenamefont
  {Noguchi}, \citenamefont {Kono}, \citenamefont {Tabuchi}, \citenamefont
  {Yamazaki}, \citenamefont {Ueda},\ and\ \citenamefont
  {Nakamura}}]{masuyama2018information}%
  \BibitemOpen
  \bibfield  {author} {\bibinfo {author} {\bibfnamefont {Y.}~\bibnamefont
  {Masuyama}}, \bibinfo {author} {\bibfnamefont {K.}~\bibnamefont {Funo}},
  \bibinfo {author} {\bibfnamefont {Y.}~\bibnamefont {Murashita}}, \bibinfo
  {author} {\bibfnamefont {A.}~\bibnamefont {Noguchi}}, \bibinfo {author}
  {\bibfnamefont {S.}~\bibnamefont {Kono}}, \bibinfo {author} {\bibfnamefont
  {Y.}~\bibnamefont {Tabuchi}}, \bibinfo {author} {\bibfnamefont
  {R.}~\bibnamefont {Yamazaki}}, \bibinfo {author} {\bibfnamefont
  {M.}~\bibnamefont {Ueda}},\ and\ \bibinfo {author} {\bibfnamefont
  {Y.}~\bibnamefont {Nakamura}},\ }\bibfield  {title} {\bibinfo {title}
  {Information-to-work conversion by maxwell’s demon in a superconducting
  circuit quantum electrodynamical system},\ }\href@noop {} {\bibfield
  {journal} {\bibinfo  {journal} {Nat. Commun.}\ }\textbf {\bibinfo {volume}
  {9}},\ \bibinfo {pages} {1} (\bibinfo {year} {2018})}\BibitemShut {NoStop}%
\bibitem [{\citenamefont {Cottet}\ \emph {et~al.}(2017)\citenamefont {Cottet},
  \citenamefont {Jezouin}, \citenamefont {Bretheau}, \citenamefont
  {Campagne-Ibarcq}, \citenamefont {Ficheux}, \citenamefont {Anders},
  \citenamefont {Auff{\`e}ves}, \citenamefont {Azouit}, \citenamefont
  {Rouchon},\ and\ \citenamefont {Huard}}]{cottet2017observing}%
  \BibitemOpen
  \bibfield  {author} {\bibinfo {author} {\bibfnamefont {N.}~\bibnamefont
  {Cottet}}, \bibinfo {author} {\bibfnamefont {S.}~\bibnamefont {Jezouin}},
  \bibinfo {author} {\bibfnamefont {L.}~\bibnamefont {Bretheau}}, \bibinfo
  {author} {\bibfnamefont {P.}~\bibnamefont {Campagne-Ibarcq}}, \bibinfo
  {author} {\bibfnamefont {Q.}~\bibnamefont {Ficheux}}, \bibinfo {author}
  {\bibfnamefont {J.}~\bibnamefont {Anders}}, \bibinfo {author} {\bibfnamefont
  {A.}~\bibnamefont {Auff{\`e}ves}}, \bibinfo {author} {\bibfnamefont
  {R.}~\bibnamefont {Azouit}}, \bibinfo {author} {\bibfnamefont
  {P.}~\bibnamefont {Rouchon}},\ and\ \bibinfo {author} {\bibfnamefont
  {B.}~\bibnamefont {Huard}},\ }\bibfield  {title} {\bibinfo {title} {Observing
  a quantum maxwell demon at work},\ }\href@noop {} {\bibfield  {journal}
  {\bibinfo  {journal} {Proc. Nat. Acad. Sci.}\ }\textbf {\bibinfo {volume}
  {114}},\ \bibinfo {pages} {7561} (\bibinfo {year} {2017})}\BibitemShut
  {NoStop}%
\bibitem [{\citenamefont {Vidrighin}\ \emph {et~al.}(2016)\citenamefont
  {Vidrighin}, \citenamefont {Dahlsten}, \citenamefont {Barbieri},
  \citenamefont {Kim}, \citenamefont {Vedral},\ and\ \citenamefont
  {Walmsley}}]{vidrighin2016photonic}%
  \BibitemOpen
  \bibfield  {author} {\bibinfo {author} {\bibfnamefont {M.~D.}\ \bibnamefont
  {Vidrighin}}, \bibinfo {author} {\bibfnamefont {O.}~\bibnamefont {Dahlsten}},
  \bibinfo {author} {\bibfnamefont {M.}~\bibnamefont {Barbieri}}, \bibinfo
  {author} {\bibfnamefont {M.}~\bibnamefont {Kim}}, \bibinfo {author}
  {\bibfnamefont {V.}~\bibnamefont {Vedral}},\ and\ \bibinfo {author}
  {\bibfnamefont {I.~A.}\ \bibnamefont {Walmsley}},\ }\bibfield  {title}
  {\bibinfo {title} {Photonic maxwell’s demon},\ }\href@noop {} {\bibfield
  {journal} {\bibinfo  {journal} {Phys. Rev. Lett.}\ }\textbf {\bibinfo
  {volume} {116}},\ \bibinfo {pages} {050401} (\bibinfo {year}
  {2016})}\BibitemShut {NoStop}%
\bibitem [{\citenamefont {Szilard}(1929)}]{szilard1929entropieverminderung}%
  \BibitemOpen
  \bibfield  {author} {\bibinfo {author} {\bibfnamefont {L.}~\bibnamefont
  {Szilard}},\ }\bibfield  {title} {\bibinfo {title} {{\"U}ber die
  entropieverminderung in einem thermodynamischen system bei eingriffen
  intelligenter wesen},\ }\href@noop {} {\bibfield  {journal} {\bibinfo
  {journal} {Zeitschrift f{\"u}r Physik}\ }\textbf {\bibinfo {volume} {53}},\
  \bibinfo {pages} {840} (\bibinfo {year} {1929})}\BibitemShut {NoStop}%
\bibitem [{\citenamefont {Kim}\ and\ \citenamefont
  {Kim}(2011)}]{kim2011information}%
  \BibitemOpen
  \bibfield  {author} {\bibinfo {author} {\bibfnamefont {K.~H.}\ \bibnamefont
  {Kim}}\ and\ \bibinfo {author} {\bibfnamefont {S.~W.}\ \bibnamefont {Kim}},\
  }\bibfield  {title} {\bibinfo {title} {Information from time-forward and
  time-backward processes in szilard engines},\ }\href@noop {} {\bibfield
  {journal} {\bibinfo  {journal} {Phys. Rev. E}\ }\textbf {\bibinfo {volume}
  {84}},\ \bibinfo {pages} {012101} (\bibinfo {year} {2011})}\BibitemShut
  {NoStop}%
\bibitem [{\citenamefont {Horowitz}\ and\ \citenamefont
  {Parrondo}(2011)}]{horowitz2011designing}%
  \BibitemOpen
  \bibfield  {author} {\bibinfo {author} {\bibfnamefont {J.~M.}\ \bibnamefont
  {Horowitz}}\ and\ \bibinfo {author} {\bibfnamefont {J.~M.}\ \bibnamefont
  {Parrondo}},\ }\bibfield  {title} {\bibinfo {title} {Designing optimal
  discrete-feedback thermodynamic engines},\ }\href@noop {} {\bibfield
  {journal} {\bibinfo  {journal} {New J. Phys.}\ }\textbf {\bibinfo {volume}
  {13}},\ \bibinfo {pages} {123019} (\bibinfo {year} {2011})}\BibitemShut
  {NoStop}%
\bibitem [{\citenamefont {Song}\ \emph {et~al.}(2021)\citenamefont {Song},
  \citenamefont {Still}, \citenamefont {Rojas}, \citenamefont {Castillo},\ and\
  \citenamefont {Marsili}}]{song2021optimal}%
  \BibitemOpen
  \bibfield  {author} {\bibinfo {author} {\bibfnamefont {J.}~\bibnamefont
  {Song}}, \bibinfo {author} {\bibfnamefont {S.}~\bibnamefont {Still}},
  \bibinfo {author} {\bibfnamefont {R.~D.~H.}\ \bibnamefont {Rojas}}, \bibinfo
  {author} {\bibfnamefont {I.~P.}\ \bibnamefont {Castillo}},\ and\ \bibinfo
  {author} {\bibfnamefont {M.}~\bibnamefont {Marsili}},\ }\bibfield  {title}
  {\bibinfo {title} {Optimal work extraction and mutual information in a
  generalized szil{\'a}rd engine},\ }\href@noop {} {\bibfield  {journal}
  {\bibinfo  {journal} {Phys. Rev. E}\ }\textbf {\bibinfo {volume} {103}},\
  \bibinfo {pages} {052121} (\bibinfo {year} {2021})}\BibitemShut {NoStop}%
\bibitem [{\citenamefont {Kim}\ \emph {et~al.}(2011)\citenamefont {Kim},
  \citenamefont {Sagawa}, \citenamefont {De~Liberato},\ and\ \citenamefont
  {Ueda}}]{kim2011quantum}%
  \BibitemOpen
  \bibfield  {author} {\bibinfo {author} {\bibfnamefont {S.~W.}\ \bibnamefont
  {Kim}}, \bibinfo {author} {\bibfnamefont {T.}~\bibnamefont {Sagawa}},
  \bibinfo {author} {\bibfnamefont {S.}~\bibnamefont {De~Liberato}},\ and\
  \bibinfo {author} {\bibfnamefont {M.}~\bibnamefont {Ueda}},\ }\bibfield
  {title} {\bibinfo {title} {Quantum szilard engine},\ }\href@noop {}
  {\bibfield  {journal} {\bibinfo  {journal} {Phys. Rev. Lett.}\ }\textbf
  {\bibinfo {volume} {106}},\ \bibinfo {pages} {070401} (\bibinfo {year}
  {2011})}\BibitemShut {NoStop}%
\bibitem [{\citenamefont {Bengtsson}\ \emph {et~al.}(2018)\citenamefont
  {Bengtsson}, \citenamefont {Tengstrand}, \citenamefont {Wacker},
  \citenamefont {Samuelsson}, \citenamefont {Ueda}, \citenamefont {Linke},\
  and\ \citenamefont {Reimann}}]{bengtsson2018quantum}%
  \BibitemOpen
  \bibfield  {author} {\bibinfo {author} {\bibfnamefont {J.}~\bibnamefont
  {Bengtsson}}, \bibinfo {author} {\bibfnamefont {M.~N.}\ \bibnamefont
  {Tengstrand}}, \bibinfo {author} {\bibfnamefont {A.}~\bibnamefont {Wacker}},
  \bibinfo {author} {\bibfnamefont {P.}~\bibnamefont {Samuelsson}}, \bibinfo
  {author} {\bibfnamefont {M.}~\bibnamefont {Ueda}}, \bibinfo {author}
  {\bibfnamefont {H.}~\bibnamefont {Linke}},\ and\ \bibinfo {author}
  {\bibfnamefont {S.}~\bibnamefont {Reimann}},\ }\bibfield  {title} {\bibinfo
  {title} {Quantum szilard engine with attractively interacting bosons},\
  }\href@noop {} {\bibfield  {journal} {\bibinfo  {journal} {Phys. Rev. Lett.}\
  }\textbf {\bibinfo {volume} {120}},\ \bibinfo {pages} {100601} (\bibinfo
  {year} {2018})}\BibitemShut {NoStop}%
\bibitem [{\citenamefont {Cai}\ \emph {et~al.}(2012)\citenamefont {Cai},
  \citenamefont {Dong}, \citenamefont {Sun} \emph
  {et~al.}}]{cai2012multiparticle}%
  \BibitemOpen
  \bibfield  {author} {\bibinfo {author} {\bibfnamefont {C.}~\bibnamefont
  {Cai}}, \bibinfo {author} {\bibfnamefont {H.}~\bibnamefont {Dong}}, \bibinfo
  {author} {\bibfnamefont {C.}~\bibnamefont {Sun}}, \emph {et~al.},\ }\bibfield
   {title} {\bibinfo {title} {Multiparticle quantum szilard engine with optimal
  cycles assisted by a maxwell's demon},\ }\href@noop {} {\bibfield  {journal}
  {\bibinfo  {journal} {Phys. Rev. E}\ }\textbf {\bibinfo {volume} {85}},\
  \bibinfo {pages} {031114} (\bibinfo {year} {2012})}\BibitemShut {NoStop}%
\bibitem [{\citenamefont {Jeon}\ and\ \citenamefont
  {Kim}(2016)}]{jeon2016optimal}%
  \BibitemOpen
  \bibfield  {author} {\bibinfo {author} {\bibfnamefont {H.~J.}\ \bibnamefont
  {Jeon}}\ and\ \bibinfo {author} {\bibfnamefont {S.~W.}\ \bibnamefont {Kim}},\
  }\bibfield  {title} {\bibinfo {title} {Optimal work of the quantum szilard
  engine under isothermal processes with inevitable irreversibility},\
  }\href@noop {} {\bibfield  {journal} {\bibinfo  {journal} {New J. Phys.}\
  }\textbf {\bibinfo {volume} {18}},\ \bibinfo {pages} {043002} (\bibinfo
  {year} {2016})}\BibitemShut {NoStop}%
\bibitem [{com()}]{comment0}%
  \BibitemOpen
  \href@noop {} {\bibinfo  {journal} {See Supplemental Material at [URL will be
  inserted by publisher].}\ }\BibitemShut {NoStop}%
\bibitem [{\citenamefont {Saha}\ \emph {et~al.}(2023)\citenamefont {Saha},
  \citenamefont {Ehrich}, \citenamefont {Gavrilov}, \citenamefont {Still},
  \citenamefont {Sivak},\ and\ \citenamefont
  {Bechhoefer}}]{saha2023information}%
  \BibitemOpen
\bibfield  {journal} {  }\bibfield  {author} {\bibinfo {author} {\bibfnamefont
  {T.~K.}\ \bibnamefont {Saha}}, \bibinfo {author} {\bibfnamefont
  {J.}~\bibnamefont {Ehrich}}, \bibinfo {author} {\bibfnamefont
  {M.}~\bibnamefont {Gavrilov}}, \bibinfo {author} {\bibfnamefont
  {S.}~\bibnamefont {Still}}, \bibinfo {author} {\bibfnamefont {D.~A.}\
  \bibnamefont {Sivak}},\ and\ \bibinfo {author} {\bibfnamefont
  {J.}~\bibnamefont {Bechhoefer}},\ }\bibfield  {title} {\bibinfo {title}
  {Information engine in a nonequilibrium bath},\ }\href@noop {} {\bibfield
  {journal} {\bibinfo  {journal} {Phys. Rev. Lett.}\ }\textbf {\bibinfo
  {volume} {131}},\ \bibinfo {pages} {057101} (\bibinfo {year}
  {2023})}\BibitemShut {NoStop}%
\bibitem [{\citenamefont {Paneru}\ \emph {et~al.}(2022)\citenamefont {Paneru},
  \citenamefont {Dutta},\ and\ \citenamefont {Pak}}]{paneru2022colossal}%
  \BibitemOpen
  \bibfield  {author} {\bibinfo {author} {\bibfnamefont {G.}~\bibnamefont
  {Paneru}}, \bibinfo {author} {\bibfnamefont {S.}~\bibnamefont {Dutta}},\ and\
  \bibinfo {author} {\bibfnamefont {H.~K.}\ \bibnamefont {Pak}},\ }\bibfield
  {title} {\bibinfo {title} {Colossal power extraction from active cyclic
  brownian information engines},\ }\href@noop {} {\bibfield  {journal}
  {\bibinfo  {journal} {J. Phys. Chem. Lett.}\ }\textbf {\bibinfo {volume}
  {13}},\ \bibinfo {pages} {6912} (\bibinfo {year} {2022})}\BibitemShut
  {NoStop}%
\bibitem [{\citenamefont {Malgaretti}\ and\ \citenamefont
  {Stark}(2022)}]{malgaretti2022szilard}%
  \BibitemOpen
  \bibfield  {author} {\bibinfo {author} {\bibfnamefont {P.}~\bibnamefont
  {Malgaretti}}\ and\ \bibinfo {author} {\bibfnamefont {H.}~\bibnamefont
  {Stark}},\ }\bibfield  {title} {\bibinfo {title} {Szilard engines and
  information-based work extraction for active systems},\ }\href@noop {}
  {\bibfield  {journal} {\bibinfo  {journal} {Phys. Rev. Lett.}\ }\textbf
  {\bibinfo {volume} {129}},\ \bibinfo {pages} {228005} (\bibinfo {year}
  {2022})}\BibitemShut {NoStop}%
\bibitem [{\citenamefont {Goldhirsch}\ and\ \citenamefont
  {Zanetti}(1993)}]{Goldhirsch93}%
  \BibitemOpen
  \bibfield  {author} {\bibinfo {author} {\bibfnamefont {I.}~\bibnamefont
  {Goldhirsch}}\ and\ \bibinfo {author} {\bibfnamefont {G.}~\bibnamefont
  {Zanetti}},\ }\bibfield  {title} {\bibinfo {title} {Clustering instability in
  dissipative gases},\ }\href {https://doi.org/10.1103/PhysRevLett.70.1619}
  {\bibfield  {journal} {\bibinfo  {journal} {Phys. Rev. Lett.}\ }\textbf
  {\bibinfo {volume} {70}},\ \bibinfo {pages} {1619} (\bibinfo {year}
  {1993})}\BibitemShut {NoStop}%
\bibitem [{\citenamefont {Narayan}\ \emph {et~al.}(2007)\citenamefont
  {Narayan}, \citenamefont {Ramaswamy},\ and\ \citenamefont
  {Menon}}]{Narayan2007}%
  \BibitemOpen
  \bibfield  {author} {\bibinfo {author} {\bibfnamefont {V.}~\bibnamefont
  {Narayan}}, \bibinfo {author} {\bibfnamefont {S.}~\bibnamefont {Ramaswamy}},\
  and\ \bibinfo {author} {\bibfnamefont {N.}~\bibnamefont {Menon}},\ }\bibfield
   {title} {\bibinfo {title} {Long-lived giant number fluctuations in a
  swarming granular nematic},\ }\href@noop {} {\bibfield  {journal} {\bibinfo
  {journal} {Science}\ }\textbf {\bibinfo {volume} {317}},\ \bibinfo {pages}
  {105} (\bibinfo {year} {2007})}\BibitemShut {NoStop}%
\bibitem [{\citenamefont {Deseigne}\ \emph {et~al.}(2010)\citenamefont
  {Deseigne}, \citenamefont {Dauchot},\ and\ \citenamefont
  {Chat{\'e}}}]{deseigne2010collective}%
  \BibitemOpen
  \bibfield  {author} {\bibinfo {author} {\bibfnamefont {J.}~\bibnamefont
  {Deseigne}}, \bibinfo {author} {\bibfnamefont {O.}~\bibnamefont {Dauchot}},\
  and\ \bibinfo {author} {\bibfnamefont {H.}~\bibnamefont {Chat{\'e}}},\
  }\bibfield  {title} {\bibinfo {title} {Collective motion of vibrated polar
  disks},\ }\href@noop {} {\bibfield  {journal} {\bibinfo  {journal} {Phys.
  Rev. Lett.}\ }\textbf {\bibinfo {volume} {105}},\ \bibinfo {pages} {098001}
  (\bibinfo {year} {2010})}\BibitemShut {NoStop}%
\bibitem [{\citenamefont {Ramaswamy}\ \emph {et~al.}(2003)\citenamefont
  {Ramaswamy}, \citenamefont {Simha},\ and\ \citenamefont
  {Toner}}]{Ramaswamy2003}%
  \BibitemOpen
  \bibfield  {author} {\bibinfo {author} {\bibfnamefont {S.}~\bibnamefont
  {Ramaswamy}}, \bibinfo {author} {\bibfnamefont {R.~A.}\ \bibnamefont
  {Simha}},\ and\ \bibinfo {author} {\bibfnamefont {J.}~\bibnamefont {Toner}},\
  }\bibfield  {title} {\bibinfo {title} {Active nematics on a substrate: Giant
  number fluctuations and long-time tails},\ }\href
  {https://doi.org/10.1209/epl/i2003-00346-7} {\bibfield  {journal} {\bibinfo
  {journal} {Europhys. Lett.}\ }\textbf {\bibinfo {volume} {62}},\ \bibinfo
  {pages} {196} (\bibinfo {year} {2003})}\BibitemShut {NoStop}%
\bibitem [{\citenamefont {Toner}\ \emph {et~al.}(2005)\citenamefont {Toner},
  \citenamefont {Tu},\ and\ \citenamefont
  {Ramaswamy}}]{toner2005hydrodynamics}%
  \BibitemOpen
  \bibfield  {author} {\bibinfo {author} {\bibfnamefont {J.}~\bibnamefont
  {Toner}}, \bibinfo {author} {\bibfnamefont {Y.}~\bibnamefont {Tu}},\ and\
  \bibinfo {author} {\bibfnamefont {S.}~\bibnamefont {Ramaswamy}},\ }\bibfield
  {title} {\bibinfo {title} {Hydrodynamics and phases of flocks},\ }\href@noop
  {} {\bibfield  {journal} {\bibinfo  {journal} {Ann. Phys.}\ }\textbf
  {\bibinfo {volume} {318}},\ \bibinfo {pages} {170} (\bibinfo {year}
  {2005})}\BibitemShut {NoStop}%
\bibitem [{\citenamefont {Ginelli}\ \emph {et~al.}(2010)\citenamefont
  {Ginelli}, \citenamefont {Peruani}, \citenamefont {B{\"a}r},\ and\
  \citenamefont {Chat{\'e}}}]{ginelli2010large}%
  \BibitemOpen
  \bibfield  {author} {\bibinfo {author} {\bibfnamefont {F.}~\bibnamefont
  {Ginelli}}, \bibinfo {author} {\bibfnamefont {F.}~\bibnamefont {Peruani}},
  \bibinfo {author} {\bibfnamefont {M.}~\bibnamefont {B{\"a}r}},\ and\ \bibinfo
  {author} {\bibfnamefont {H.}~\bibnamefont {Chat{\'e}}},\ }\bibfield  {title}
  {\bibinfo {title} {Large-scale collective properties of self-propelled
  rods},\ }\href@noop {} {\bibfield  {journal} {\bibinfo  {journal} {Phys. Rev.
  Lett.}\ }\textbf {\bibinfo {volume} {104}},\ \bibinfo {pages} {184502}
  (\bibinfo {year} {2010})}\BibitemShut {NoStop}%
\bibitem [{\citenamefont {Chat{\'e}}\ \emph {et~al.}(2008)\citenamefont
  {Chat{\'e}}, \citenamefont {Ginelli}, \citenamefont {Gr{\'e}goire},\ and\
  \citenamefont {Raynaud}}]{chate2008collective}%
  \BibitemOpen
  \bibfield  {author} {\bibinfo {author} {\bibfnamefont {H.}~\bibnamefont
  {Chat{\'e}}}, \bibinfo {author} {\bibfnamefont {F.}~\bibnamefont {Ginelli}},
  \bibinfo {author} {\bibfnamefont {G.}~\bibnamefont {Gr{\'e}goire}},\ and\
  \bibinfo {author} {\bibfnamefont {F.}~\bibnamefont {Raynaud}},\ }\bibfield
  {title} {\bibinfo {title} {Collective motion of self-propelled particles
  interacting without cohesion},\ }\href@noop {} {\bibfield  {journal}
  {\bibinfo  {journal} {Phys. Rev. E}\ }\textbf {\bibinfo {volume} {77}},\
  \bibinfo {pages} {046113} (\bibinfo {year} {2008})}\BibitemShut {NoStop}%
\bibitem [{\citenamefont {Zhang}\ \emph {et~al.}(2010)\citenamefont {Zhang},
  \citenamefont {Be’er}, \citenamefont {Florin},\ and\ \citenamefont
  {Swinney}}]{zhang2010collective}%
  \BibitemOpen
  \bibfield  {author} {\bibinfo {author} {\bibfnamefont {H.-P.}\ \bibnamefont
  {Zhang}}, \bibinfo {author} {\bibfnamefont {A.}~\bibnamefont {Be’er}},
  \bibinfo {author} {\bibfnamefont {E.-L.}\ \bibnamefont {Florin}},\ and\
  \bibinfo {author} {\bibfnamefont {H.~L.}\ \bibnamefont {Swinney}},\
  }\bibfield  {title} {\bibinfo {title} {Collective motion and density
  fluctuations in bacterial colonies},\ }\href@noop {} {\bibfield  {journal}
  {\bibinfo  {journal} {Proc. Nat. Acad. Sci.}\ }\textbf {\bibinfo {volume}
  {107}},\ \bibinfo {pages} {13626} (\bibinfo {year} {2010})}\BibitemShut
  {NoStop}%
\bibitem [{\citenamefont {Palacci}\ \emph {et~al.}(2013)\citenamefont
  {Palacci}, \citenamefont {Sacanna}, \citenamefont {Steinberg}, \citenamefont
  {Pine},\ and\ \citenamefont {Chaikin}}]{Palacci2013}%
  \BibitemOpen
  \bibfield  {author} {\bibinfo {author} {\bibfnamefont {J.}~\bibnamefont
  {Palacci}}, \bibinfo {author} {\bibfnamefont {S.}~\bibnamefont {Sacanna}},
  \bibinfo {author} {\bibfnamefont {A.~P.}\ \bibnamefont {Steinberg}}, \bibinfo
  {author} {\bibfnamefont {D.~J.}\ \bibnamefont {Pine}},\ and\ \bibinfo
  {author} {\bibfnamefont {P.~M.}\ \bibnamefont {Chaikin}},\ }\bibfield
  {title} {\bibinfo {title} {Living crystals of light-activated colloidal
  surfers},\ }\href@noop {} {\bibfield  {journal} {\bibinfo  {journal}
  {Science}\ }\textbf {\bibinfo {volume} {339}},\ \bibinfo {pages} {936}
  (\bibinfo {year} {2013})}\BibitemShut {NoStop}%
\bibitem [{\citenamefont {Bricard}\ \emph {et~al.}(2013)\citenamefont
  {Bricard}, \citenamefont {Caussin}, \citenamefont {Desreumaux}, \citenamefont
  {Dauchot},\ and\ \citenamefont {Bartolo}}]{bricard2013emergence}%
  \BibitemOpen
  \bibfield  {author} {\bibinfo {author} {\bibfnamefont {A.}~\bibnamefont
  {Bricard}}, \bibinfo {author} {\bibfnamefont {J.-B.}\ \bibnamefont
  {Caussin}}, \bibinfo {author} {\bibfnamefont {N.}~\bibnamefont {Desreumaux}},
  \bibinfo {author} {\bibfnamefont {O.}~\bibnamefont {Dauchot}},\ and\ \bibinfo
  {author} {\bibfnamefont {D.}~\bibnamefont {Bartolo}},\ }\bibfield  {title}
  {\bibinfo {title} {Emergence of macroscopic directed motion in populations of
  motile colloids},\ }\href@noop {} {\bibfield  {journal} {\bibinfo  {journal}
  {Nature}\ }\textbf {\bibinfo {volume} {503}},\ \bibinfo {pages} {95}
  (\bibinfo {year} {2013})}\BibitemShut {NoStop}%
\bibitem [{\citenamefont {Giomi}\ \emph {et~al.}(2013)\citenamefont {Giomi},
  \citenamefont {Hawley-Weld},\ and\ \citenamefont
  {Mahadevan}}]{giomi2013swarming}%
  \BibitemOpen
  \bibfield  {author} {\bibinfo {author} {\bibfnamefont {L.}~\bibnamefont
  {Giomi}}, \bibinfo {author} {\bibfnamefont {N.}~\bibnamefont {Hawley-Weld}},\
  and\ \bibinfo {author} {\bibfnamefont {L.}~\bibnamefont {Mahadevan}},\
  }\bibfield  {title} {\bibinfo {title} {Swarming, swirling and stasis in
  sequestered bristle-bots},\ }\href@noop {} {\bibfield  {journal} {\bibinfo
  {journal} {Proc. R. Soc. London, Ser. A}\ }\textbf {\bibinfo {volume}
  {469}},\ \bibinfo {pages} {20120637} (\bibinfo {year} {2013})}\BibitemShut
  {NoStop}%
\bibitem [{\citenamefont {Deblais}\ \emph {et~al.}(2018)\citenamefont
  {Deblais}, \citenamefont {Barois}, \citenamefont {Guerin}, \citenamefont
  {Delville}, \citenamefont {Vaudaine}, \citenamefont {Lintuvuori},
  \citenamefont {Boudet}, \citenamefont {Baret},\ and\ \citenamefont
  {Kellay}}]{deblais2018boundaries}%
  \BibitemOpen
  \bibfield  {author} {\bibinfo {author} {\bibfnamefont {A.}~\bibnamefont
  {Deblais}}, \bibinfo {author} {\bibfnamefont {T.}~\bibnamefont {Barois}},
  \bibinfo {author} {\bibfnamefont {T.}~\bibnamefont {Guerin}}, \bibinfo
  {author} {\bibfnamefont {P.-H.}\ \bibnamefont {Delville}}, \bibinfo {author}
  {\bibfnamefont {R.}~\bibnamefont {Vaudaine}}, \bibinfo {author}
  {\bibfnamefont {J.~S.}\ \bibnamefont {Lintuvuori}}, \bibinfo {author}
  {\bibfnamefont {J.-F.}\ \bibnamefont {Boudet}}, \bibinfo {author}
  {\bibfnamefont {J.-C.}\ \bibnamefont {Baret}},\ and\ \bibinfo {author}
  {\bibfnamefont {H.}~\bibnamefont {Kellay}},\ }\bibfield  {title} {\bibinfo
  {title} {Boundaries control collective dynamics of inertial self-propelled
  robots},\ }\href@noop {} {\bibfield  {journal} {\bibinfo  {journal} {Phys.
  Rev. Lett.}\ }\textbf {\bibinfo {volume} {120}},\ \bibinfo {pages} {188002}
  (\bibinfo {year} {2018})}\BibitemShut {NoStop}%
\bibitem [{\citenamefont {S{\'a}nchez}\ and\ \citenamefont
  {D{\'\i}az-Leyva}(2018)}]{sanchez2018self}%
  \BibitemOpen
  \bibfield  {author} {\bibinfo {author} {\bibfnamefont {R.}~\bibnamefont
  {S{\'a}nchez}}\ and\ \bibinfo {author} {\bibfnamefont {P.}~\bibnamefont
  {D{\'\i}az-Leyva}},\ }\bibfield  {title} {\bibinfo {title} {Self-assembly and
  speed distributions of active granular particles},\ }\href@noop {} {\bibfield
   {journal} {\bibinfo  {journal} {Phys. A}\ }\textbf {\bibinfo {volume}
  {499}},\ \bibinfo {pages} {11} (\bibinfo {year} {2018})}\BibitemShut
  {NoStop}%
\bibitem [{\citenamefont {Dauchot}\ and\ \citenamefont
  {D{\'e}mery}(2019)}]{dauchot2019dynamics}%
  \BibitemOpen
  \bibfield  {author} {\bibinfo {author} {\bibfnamefont {O.}~\bibnamefont
  {Dauchot}}\ and\ \bibinfo {author} {\bibfnamefont {V.}~\bibnamefont
  {D{\'e}mery}},\ }\bibfield  {title} {\bibinfo {title} {Dynamics of a
  self-propelled particle in a harmonic trap},\ }\href@noop {} {\bibfield
  {journal} {\bibinfo  {journal} {Phys. Rev. Lett.}\ }\textbf {\bibinfo
  {volume} {122}},\ \bibinfo {pages} {068002} (\bibinfo {year}
  {2019})}\BibitemShut {NoStop}%
\bibitem [{\citenamefont {Bradski}(2000)}]{bradski2000opencv}%
  \BibitemOpen
  \bibfield  {author} {\bibinfo {author} {\bibfnamefont {G.}~\bibnamefont
  {Bradski}},\ }\bibfield  {title} {\bibinfo {title} {The opencv library.},\
  }\href@noop {} {\bibfield  {journal} {\bibinfo  {journal} {Dr. Dobb's
  Journal: Software Tools for the Professional Programmer}\ }\textbf {\bibinfo
  {volume} {25}},\ \bibinfo {pages} {120} (\bibinfo {year} {2000})}\BibitemShut
  {NoStop}%
\bibitem [{\citenamefont {Park}\ \emph {et~al.}(2016)\citenamefont {Park},
  \citenamefont {Lee},\ and\ \citenamefont {Noh}}]{park2016optimal}%
  \BibitemOpen
  \bibfield  {author} {\bibinfo {author} {\bibfnamefont {J.-M.}\ \bibnamefont
  {Park}}, \bibinfo {author} {\bibfnamefont {J.~S.}\ \bibnamefont {Lee}},\ and\
  \bibinfo {author} {\bibfnamefont {J.~D.}\ \bibnamefont {Noh}},\ }\bibfield
  {title} {\bibinfo {title} {Optimal tuning of a confined brownian information
  engine},\ }\href@noop {} {\bibfield  {journal} {\bibinfo  {journal} {Phys.
  Rev. E}\ }\textbf {\bibinfo {volume} {93}},\ \bibinfo {pages} {032146}
  (\bibinfo {year} {2016})}\BibitemShut {NoStop}%
\bibitem [{\citenamefont {Andresen}\ \emph {et~al.}(1977)\citenamefont
  {Andresen}, \citenamefont {Berry}, \citenamefont {Nitzan},\ and\
  \citenamefont {Salamon}}]{andresen1977thermodynamics}%
  \BibitemOpen
  \bibfield  {author} {\bibinfo {author} {\bibfnamefont {B.}~\bibnamefont
  {Andresen}}, \bibinfo {author} {\bibfnamefont {R.~S.}\ \bibnamefont {Berry}},
  \bibinfo {author} {\bibfnamefont {A.}~\bibnamefont {Nitzan}},\ and\ \bibinfo
  {author} {\bibfnamefont {P.}~\bibnamefont {Salamon}},\ }\bibfield  {title}
  {\bibinfo {title} {Thermodynamics in finite time. i. the step-carnot cycle},\
  }\href@noop {} {\bibfield  {journal} {\bibinfo  {journal} {Phys. Rev. A}\
  }\textbf {\bibinfo {volume} {15}},\ \bibinfo {pages} {2086} (\bibinfo {year}
  {1977})}\BibitemShut {NoStop}%
\bibitem [{\citenamefont {Hiratsuka}\ \emph {et~al.}(2006)\citenamefont
  {Hiratsuka}, \citenamefont {Miyata}, \citenamefont {Tada},\ and\
  \citenamefont {Uyeda}}]{hiratsuka2006microrotary}%
  \BibitemOpen
  \bibfield  {author} {\bibinfo {author} {\bibfnamefont {Y.}~\bibnamefont
  {Hiratsuka}}, \bibinfo {author} {\bibfnamefont {M.}~\bibnamefont {Miyata}},
  \bibinfo {author} {\bibfnamefont {T.}~\bibnamefont {Tada}},\ and\ \bibinfo
  {author} {\bibfnamefont {T.~Q.}\ \bibnamefont {Uyeda}},\ }\bibfield  {title}
  {\bibinfo {title} {A microrotary motor powered by bacteria},\ }\href@noop {}
  {\bibfield  {journal} {\bibinfo  {journal} {Proc. Nat. Acad. Sci.}\ }\textbf
  {\bibinfo {volume} {103}},\ \bibinfo {pages} {13618} (\bibinfo {year}
  {2006})}\BibitemShut {NoStop}%
\bibitem [{\citenamefont {Angelani}\ \emph {et~al.}(2009)\citenamefont
  {Angelani}, \citenamefont {Di~Leonardo},\ and\ \citenamefont
  {Ruocco}}]{angelani2009self}%
  \BibitemOpen
  \bibfield  {author} {\bibinfo {author} {\bibfnamefont {L.}~\bibnamefont
  {Angelani}}, \bibinfo {author} {\bibfnamefont {R.}~\bibnamefont
  {Di~Leonardo}},\ and\ \bibinfo {author} {\bibfnamefont {G.}~\bibnamefont
  {Ruocco}},\ }\bibfield  {title} {\bibinfo {title} {Self-starting micromotors
  in a bacterial bath},\ }\href@noop {} {\bibfield  {journal} {\bibinfo
  {journal} {Phys. Rev. Lett.}\ }\textbf {\bibinfo {volume} {102}},\ \bibinfo
  {pages} {048104} (\bibinfo {year} {2009})}\BibitemShut {NoStop}%
\bibitem [{\citenamefont {Di~Leonardo}\ \emph {et~al.}(2010)\citenamefont
  {Di~Leonardo}, \citenamefont {Angelani}, \citenamefont {Dell’Arciprete},
  \citenamefont {Ruocco}, \citenamefont {Iebba}, \citenamefont {Schippa},
  \citenamefont {Conte}, \citenamefont {Mecarini}, \citenamefont {De~Angelis},\
  and\ \citenamefont {Di~Fabrizio}}]{di2010bacterial}%
  \BibitemOpen
  \bibfield  {author} {\bibinfo {author} {\bibfnamefont {R.}~\bibnamefont
  {Di~Leonardo}}, \bibinfo {author} {\bibfnamefont {L.}~\bibnamefont
  {Angelani}}, \bibinfo {author} {\bibfnamefont {D.}~\bibnamefont
  {Dell’Arciprete}}, \bibinfo {author} {\bibfnamefont {G.}~\bibnamefont
  {Ruocco}}, \bibinfo {author} {\bibfnamefont {V.}~\bibnamefont {Iebba}},
  \bibinfo {author} {\bibfnamefont {S.}~\bibnamefont {Schippa}}, \bibinfo
  {author} {\bibfnamefont {M.~P.}\ \bibnamefont {Conte}}, \bibinfo {author}
  {\bibfnamefont {F.}~\bibnamefont {Mecarini}}, \bibinfo {author}
  {\bibfnamefont {F.}~\bibnamefont {De~Angelis}},\ and\ \bibinfo {author}
  {\bibfnamefont {E.}~\bibnamefont {Di~Fabrizio}},\ }\bibfield  {title}
  {\bibinfo {title} {Bacterial ratchet motors},\ }\href@noop {} {\bibfield
  {journal} {\bibinfo  {journal} {Proc. Nat. Acad. Sci.}\ }\textbf {\bibinfo
  {volume} {107}},\ \bibinfo {pages} {9541} (\bibinfo {year}
  {2010})}\BibitemShut {NoStop}%
\bibitem [{\citenamefont {Angelani}\ \emph {et~al.}(2011)\citenamefont
  {Angelani}, \citenamefont {Costanzo},\ and\ \citenamefont
  {Di~Leonardo}}]{angelani2011active}%
  \BibitemOpen
  \bibfield  {author} {\bibinfo {author} {\bibfnamefont {L.}~\bibnamefont
  {Angelani}}, \bibinfo {author} {\bibfnamefont {A.}~\bibnamefont {Costanzo}},\
  and\ \bibinfo {author} {\bibfnamefont {R.}~\bibnamefont {Di~Leonardo}},\
  }\bibfield  {title} {\bibinfo {title} {Active ratchets},\ }\href@noop {}
  {\bibfield  {journal} {\bibinfo  {journal} {Europhys. Lett.}\ }\textbf
  {\bibinfo {volume} {96}},\ \bibinfo {pages} {68002} (\bibinfo {year}
  {2011})}\BibitemShut {NoStop}%
\bibitem [{\citenamefont {Lagoin}\ \emph {et~al.}(2022)\citenamefont {Lagoin},
  \citenamefont {Crauste-Thibierge},\ and\ \citenamefont
  {Naert}}]{lagoin2022human}%
  \BibitemOpen
  \bibfield  {author} {\bibinfo {author} {\bibfnamefont {M.}~\bibnamefont
  {Lagoin}}, \bibinfo {author} {\bibfnamefont {C.}~\bibnamefont
  {Crauste-Thibierge}},\ and\ \bibinfo {author} {\bibfnamefont
  {A.}~\bibnamefont {Naert}},\ }\bibfield  {title} {\bibinfo {title}
  {Human-scale brownian ratchet: A historical thought experiment},\ }\href@noop
  {} {\bibfield  {journal} {\bibinfo  {journal} {Phys. Rev. Lett.}\ }\textbf
  {\bibinfo {volume} {129}},\ \bibinfo {pages} {120606} (\bibinfo {year}
  {2022})}\BibitemShut {NoStop}%
\bibitem [{\citenamefont {Gupta}\ \emph {et~al.}(2023)\citenamefont {Gupta},
  \citenamefont {Klapp},\ and\ \citenamefont {Sivak}}]{gupta2023efficient}%
  \BibitemOpen
  \bibfield  {author} {\bibinfo {author} {\bibfnamefont {D.}~\bibnamefont
  {Gupta}}, \bibinfo {author} {\bibfnamefont {S.~H.}\ \bibnamefont {Klapp}},\
  and\ \bibinfo {author} {\bibfnamefont {D.~A.}\ \bibnamefont {Sivak}},\
  }\bibfield  {title} {\bibinfo {title} {Efficient control protocols for an
  active ornstein-uhlenbeck particle},\ }\href@noop {} {\bibfield  {journal}
  {\bibinfo  {journal} {arXiv preprint arXiv:2304.12926}\ } (\bibinfo {year}
  {2023})}\BibitemShut {NoStop}%
\bibitem [{\citenamefont {Norton}\ \emph {et~al.}(2020)\citenamefont {Norton},
  \citenamefont {Grover}, \citenamefont {Hagan},\ and\ \citenamefont
  {Fraden}}]{norton2020optimal}%
  \BibitemOpen
  \bibfield  {author} {\bibinfo {author} {\bibfnamefont {M.~M.}\ \bibnamefont
  {Norton}}, \bibinfo {author} {\bibfnamefont {P.}~\bibnamefont {Grover}},
  \bibinfo {author} {\bibfnamefont {M.~F.}\ \bibnamefont {Hagan}},\ and\
  \bibinfo {author} {\bibfnamefont {S.}~\bibnamefont {Fraden}},\ }\bibfield
  {title} {\bibinfo {title} {Optimal control of active nematics},\ }\href@noop
  {} {\bibfield  {journal} {\bibinfo  {journal} {Phys. Rev. Lett.}\ }\textbf
  {\bibinfo {volume} {125}},\ \bibinfo {pages} {178005} (\bibinfo {year}
  {2020})}\BibitemShut {NoStop}%
\bibitem [{\citenamefont {Shankar}\ \emph {et~al.}(2022)\citenamefont
  {Shankar}, \citenamefont {Raju},\ and\ \citenamefont
  {Mahadevan}}]{shankar2022optimal}%
  \BibitemOpen
  \bibfield  {author} {\bibinfo {author} {\bibfnamefont {S.}~\bibnamefont
  {Shankar}}, \bibinfo {author} {\bibfnamefont {V.}~\bibnamefont {Raju}},\ and\
  \bibinfo {author} {\bibfnamefont {L.}~\bibnamefont {Mahadevan}},\ }\bibfield
  {title} {\bibinfo {title} {Optimal transport and control of active drops},\
  }\href@noop {} {\bibfield  {journal} {\bibinfo  {journal} {Proc. Natl. Acad.
  Sci. USA}\ }\textbf {\bibinfo {volume} {119}},\ \bibinfo {pages}
  {e2121985119} (\bibinfo {year} {2022})}\BibitemShut {NoStop}%
\bibitem [{\citenamefont {Yang}\ and\ \citenamefont
  {Bevan}(2018)}]{yang2018optimal}%
  \BibitemOpen
  \bibfield  {author} {\bibinfo {author} {\bibfnamefont {Y.}~\bibnamefont
  {Yang}}\ and\ \bibinfo {author} {\bibfnamefont {M.~A.}\ \bibnamefont
  {Bevan}},\ }\bibfield  {title} {\bibinfo {title} {Optimal navigation of
  self-propelled colloids},\ }\href@noop {} {\bibfield  {journal} {\bibinfo
  {journal} {ACS nano}\ }\textbf {\bibinfo {volume} {12}},\ \bibinfo {pages}
  {10712} (\bibinfo {year} {2018})}\BibitemShut {NoStop}%
\bibitem [{\citenamefont {Freitas}\ and\ \citenamefont
  {Esposito}(2022)}]{freitas2022maxwell}%
  \BibitemOpen
  \bibfield  {author} {\bibinfo {author} {\bibfnamefont {N.}~\bibnamefont
  {Freitas}}\ and\ \bibinfo {author} {\bibfnamefont {M.}~\bibnamefont
  {Esposito}},\ }\bibfield  {title} {\bibinfo {title} {Maxwell demon that can
  work at macroscopic scales},\ }\href@noop {} {\bibfield  {journal} {\bibinfo
  {journal} {Phys. Rev. Lett.}\ }\textbf {\bibinfo {volume} {129}},\ \bibinfo
  {pages} {120602} (\bibinfo {year} {2022})}\BibitemShut {NoStop}%
\end{thebibliography}
\end{document}